\documentclass[12pt,preprint]{aastex}
\usepackage{amsmath}
\begin{document}
\title{Polarization with a 3-dimensional Mixed Magnetic Field and Its Application to GRB 170817A}
\author{Mi-Xiang Lan$^{1}$, Jin-Jun Geng$^{2,3}$, Xue-Feng Wu$^{1,4}$, and Zi-Gao Dai$^{2,3}$}
\affil{$^{1}$Purple Mountain Observatory, Chinese Academy of Sciences,
Nanjing 210008, China; mxlan@pmo.ac.cn, xfwu@pmo.ac.cn \\
$^{2}$School of Astronomy and Space Science, Nanjing University, Nanjing 210093, China; dzg@nju.edu.cn\\
$^{3}$Key Laboratory of Modern Astronomy and Astrophysics (Nanjing
University), Ministry of Education, China\\
$^{4}$School of Astronomy and Space Science, University of Science and Technology of China,
Hefei, Anhui 230026, China}

\begin{abstract}
A large-scale ordered magnetic field plays a very important role in the formation and acceleration of a gamma-ray burst (GRB) jet.
During the GRB prompt phase, some dissipation processes may happen and disturb the magnetic field, making the field become random to some extent. Therefore, a mixed magnetic field consisting of an ordered component and a random component is plausible for the early afterglow phase. Here we investigate the polarization evolution and light curve of an afterglow under such a 3-dimensional mixed magnetic field. Three kinds of ordered component (i.e., aligned, toroidal and radial) are discussed. We find that the three cases are distinguishable through jet polarization evolution. The polarization angle for a 3D mixed magnetic field with an aligned ordered component can evolve gradually but only changes abruptly by $90^\circ$ in the toroidal and radial cases. Furthermore, during the reverse shock crossing time, the polarization degree (PD) can be non-zero for the toroidal case but roughly zero for the radial case. Since an aligned component in a jet corresponds to a magnetar central engine and a toroidal component corresponds to a black hole, GRB central engines are distinguishable through polarization observations even if the magnetic field is mixed in a jet. In addition, our polarization calculation can be applied to GRB 170817A associated with GW170817. Using the recently-observed PD upper limit $12\%$ of GRB 170817A at $t=244$ days, the magnetic field strength ratio of the ordered to random components in this afterglow is constrained to be $\lesssim0.9$.
\end{abstract}


\section{Introduction}
Gamma-ray bursts (GRBs) are luminous gamma-ray transients in the universe.
After the prompt gamma-ray emission, the ejecta of the burst will interact with an inter-stellar medium (ISM) and multi-band afterglows will be generated. There are two kinds of central engine for GRBs: a black hole plus accretion disk system (Narayan, Paczy\'nski, \& Piran 1992; Woosley 1993; M\'esz\'aros \& Rees 1997; Paczy\'nski 1998) and a millisecond magnetar (Usov 1992; Duncan \& Thompson 1992; Klu\'zniak \& Ruderman 1998; Dai \& Lu 1998a,1998b; Spruit 1999; Ruderman, Tao, \& Klu\'zniak 2000; Wheeler et al. 2000; Zhang \& M{\'e}sz{\'a}ros 2001; Bucciantini et al. 2008; Metzger et al. 2011). Different central engines will power the jet through different mechanisms, e.g., the Blandford-Znajek (BZ) mechanism (Blandford \& Znajek 1977) works for the black hole plus accretion disk system and the magnetar may produce ejecta through magnetic dipole radiation.

A large-scale ordered magnetic field often plays an important role in the jet's formation and acceleration. It is possible that there will be large-scale ordered magnetic field remnants in the GRB jets. The large-scale ordered magnetic field in the jet of the black hole plus accretion disk system is very likely to be toroidal due to the BZ mechanism, while it may be aligned for a magnetar central engine (Spruit et al. 2001). For a perpendicular rotator (i.e., the magnetic axis is perpendicular to the rotational axis), a stripped wind can be formed around a magnetar and the ordered magnetic field configuration in the wind is aligned (see Figs. 1-2 in Spruit et al. 2001).
On the other hand, shocks, magnetic reconnection or turbulence may happen during the jet propagation.
These processes will disturb the magnetic field lines, making the fields become random to some extent.

Polarization, as a probe of the magnetic field configurations (MFCs) and the jet geometry, can provide additional information compared to the traditional light curves and spectra. Polarization in the random magnetic field were investigated by many authors (e.g., Sari 1999; Waxman 2003; Wu et al. 2005; Toma et al. 2009; Lan, Wu \& Dai 2016a). To achieve a relatively large polarization degree (PD), an observer should be off-axis. Polarization in a toroidal magnetic field were considered by many authors (e.g., Lyutikov et al. 2003; Toma et al. 2009 and Lan, Wu \& Dai 2016a,b). Because of the axial symmetry, if the observer is on-axis, the observed PD is zero. An observer that is slightly off-axis but still within the jet half-opening angle will see relatively large PD. Polarization in an aligned magnetic field were investigated by several authors (e.g., Granot \& K\"{o}nigl 2003, hereafter GK03, Granot 2003 and Lan, Wu \& Dai 2016a,b). In this configuration, even an on-axis observation can get a large PD. The position angle (PA) could evolve gradually for an aligned MFC (Lan, Wu \& Dai 2016a,b). In contrast, the change of PA for the random and toroidal MFCs would only be abruptly by $90^\circ$ (Sari 1999; Toma et al. 2009; Lan, Wu \& Dai 2016a).
Therefore, different evolutions of the PA can be used as a probe of MFC in the jet and even the GRB central engines (Lan, Wu \& Dai 2016a).

The random MFC considered in Sari (1999) is 3-dimensional (3D). Conventionally, the ordered MFCs are often 2-dimensional (2D) and assumed to be confined in the shock plane (GK03, Granot 2003, Lyutikov et al. 2003, Toma et al. 2009; Lan, Wu \& Dai 2016a,b). Recently, a 2D ``mixed" magnetic field (which was also assumed to be confined in the shock plane) has been considered by Lan, Wu \& Dai (2018, hereafter LWD18). They found that such a 2D ``mixed" magnetic field is essentially large-scale ordered. The PD prediction for such a 2D ``mixed" magnetic field are similar to that of the pure ordered magnetic field. However, 3D numerical simulations have shown that the PD would decrease with increasing the randomness of the magnetic field (Deng et al. 2016). Therefore, a 3D mixed magnetic field including both the ordered and random components should be considered.

During the early afterglow phase or the late-time energy injection phase, a reverse shock (RS) will propagate into the jet or the injected wind from the central engine. Initially, the MFC in the jet or the injected wind is very likely to be large-scale ordered. However, during the jet propagation, some dissipating processes would happen and disturb the magnetic field lines. Since the MFCs affect the polarization properties significantly, more realistic MFCs (rather than pure ordered MFCs) should be considered. In this paper, the polarization properties with a 3D mixed magnetic field are investigated. This paper is arranged as follows. In Section 2, we present our model. In Section 3, numerical results are exhibited. In Section 4, we apply our polarization model to GRB 170817A. Our conclusions and discussion are given in Section 5.

\section{Polarization with a 3D mixed magnetic field}
After the GRB prompt emission, its ejecta will collide with an ISM, forming an RS propagating into the ejecta and a forward shock (FS) sweeping up the ISM. Therefore, four regions are produced, separated by two shocks: (1) the unshocked ISM, (2) the forward shocked ISM, (3) the reverse shocked ejecta and (4) the unshocked ejecta. The magnetic field carried out from the central engine is very likely to be ordered (Spruit et al. 2001). Before the early afterglow phase, the internal shock, turbulence or magnetic reconnection may happen in the jet during its propagation. The magnetic field lines in the jet will be disturbed and become less ordered (Zhang \& Yan 2011; Deng et al. 2016; Troja et al. 2017). Therefore, a 3D mixed magnetic field, including both ordered and random components, is a possible case in the GRB jet during the afterglow phase.

Here the synchrotron emission is considered. Electrons with Lorentz factor $\gamma_e$ will radiate energy around the characteristic frequency $\nu'_c=eB'\langle\sin\theta'_B\rangle\gamma_e^2/2\pi m_e c$. $e$ and $m_e$ are the charge and mass of the electron, $c$ is the speed of the light, $B'$ is the strength of the total magnetic field, $\theta'_B$ is the pitch angle of the electrons. The angle bracket denotes the average over the magnetic field directions. The prime represent the quantities in the comoving frame of the jet. We then express the observed Stokes parameters from the RS region (where a 3D mixed magnetic field exists) as
\begin{equation}
F_{\nu,rs}=\frac{1+z}{4\pi D_L^2}\frac{\sqrt{3}e^3}{m_ec^2}\int_0^{\theta_j+\theta_V} d\theta \mathcal{D}^3\sin\theta B'\int_{-\Delta\phi}^{\Delta\phi} d\phi\langle\sin\theta'_B\rangle\int d\gamma_e N(\gamma_e)F(x),
\end{equation}
\begin{equation}
Q_{\nu,rs}=\frac{1+z}{4\pi D_L^2}\frac{\sqrt{3}e^3}{m_ec^2}\int_0^{\theta_j+\theta_V} d\theta \mathcal{D}^3\sin\theta B'\int_{-\Delta\phi}^{\Delta\phi} d\phi \langle\sin\theta'_B\rangle\Pi_p\cos(2\chi_p)\int d\gamma_e  N(\gamma_e)F(x),
\end{equation}
\begin{equation}
U_{\nu,rs}=\frac{1+z}{4\pi D_L^2}\frac{\sqrt{3}e^3}{m_ec^2}\int_0^{\theta_j+\theta_V} d\theta \mathcal{D}^3\sin\theta B'\int_{-\Delta\phi}^{\Delta\phi} d\phi \langle\sin\theta'_B\rangle\Pi_p\sin(2\chi_p)\int d\gamma_e  N(\gamma_e)F(x),
\end{equation}
where $z$ is the redshift of the source, $D_L$ is its luminosity distance, $\theta_j$ and $\theta_V$ are the half opening angle of the jet and the viewing angle respectively; $\mathcal{D}=1/\gamma/(1-\beta\cos\theta)$ is the Doppler factor, $\gamma$ and $\beta$ are the bulk Lorentz factor and the velocity of the jet; the limit of the $\phi$ integration $\Delta\phi$ can be found in the previous papers (Wu et al. 2005; Toma et al. 2009; Lan, Wu \& Dai 2016a); $N(\gamma_e)$ is the energy spectrum of these electrons; $F(x)$ is the dimensionless spectrum of synchrotron radiation with $x\equiv\nu'/\nu'_c$, $\nu'=\nu(1+z)/\mathcal{D}$ is the comoving frequency; $\theta$ is the angle between the line of sight (LOS) and the local velocity of the jet element, $\phi$ is the angle in the plane of sky between the projection of the jet axis and the projection of the velocity of the jet element; the integration of $\theta$ is performed on the equal arrival time surface (EATS, Sari 1998; Huang et al. 2007); $\Pi_p$ and $\chi_p$ are the local PD and PA.

To obtain PD ($\Pi_p$) and PA ($\chi_p$) in a local fluid element, as in the previous papers (LWD18; Toma et al. 2009), we establish two coordinate systems: $\hat{x}\hat{y}\hat{\beta}$ and $\hat{1}\hat{2}\hat{k}'$, where $\hat{\beta}$ is the local velocity direction with no lateral expansion, $\hat{k}'=\mathcal{D}\left\{\hat{k}+\hat{\beta}\left[(\gamma-1)\hat{\beta}\cdot\hat{k}-\gamma\beta\right]\right\}$ is the comoving direction of the wavevector (i.e., the comoving LOS) and $\hat{y}=\hat{1}\parallel\hat{\beta}\times\hat{k}'\parallel\hat{\beta}\times\hat{k}$, $\hat{k}$ is the LOS. Considering a small region where the magnetic field has a fixed direction, the polar and azimuthal angle of total magnetic field $\vec{B}'_s$ in coordinate system $\hat{1}\hat{2}\hat{k}'$ are $\theta'_B$ and $\phi'_B$. Then the PA of the synchrotron emission in this magnetic field $\vec{B}'_s$ can be expressed as $\chi'=\phi'_B-\pi/2$. In a point-like region (Sari 1999), the direction of the magnetic field is stochastic. To get the local PD and PA, the averages over the magnetic field directions are needed. Thus the local PD and PA of the synchrotron emission from a point-like region in the comoving frame can be expressed as $\Pi'_p=\sqrt{\langle Q'_p\rangle^2+\langle U'_p\rangle^2}/\langle F'_p\rangle$ and $\chi'_p=\frac{1}{2}\arctan(\langle U'_p\rangle/\langle Q'_p\rangle)$\footnote{Notice that the angle $\chi'_p$ can not be obtained with the only use of this formula, the signs of the Stokes parameters $\langle U'_p\rangle$ and $\langle Q'_p\rangle$ are also needed (LWD18).}. The flux can be expressed as $F'_p=A_0(\sin\theta'_B)^{1-m}$ with a constant $A_0$ and a spectrum index $m$ defined by $F'_{\nu'}\propto\nu'^m$. The Stokes parameters can be expressed as $Q'_p=\Pi_0\cos2\chi'F'_p$ and $U'_p=\Pi_0\sin2\chi'F'_p$. Here, we adopt $\Pi_0=0.6$. Because PD is an invariant after Lorentz transformation, we have
\begin{align}
\Pi_p=\Pi'_p&=\frac{\sqrt{\langle Q'_p\rangle^2+\langle U'_p\rangle^2}}{\langle F'_p\rangle}  \nonumber \\
     &=\Pi_0\frac{\sqrt{\langle(\sin\theta'_B)^{1-m}\cos(2\phi'_B)\rangle^2+\langle (\sin\theta'_B)^{1-m}\sin(2\phi'_B)\rangle^2}}{\langle(\sin\theta'_B)^{1-m}\rangle},
\end{align}

We next derive the local PA ($\chi_p$) in the observer frame. The averaged electric vector in the comoving frame can be expressed as $\hat{\overline{e}'}=\cos\chi'_p\hat{1}+\sin\chi'_p\hat{2}$ with $\hat{1}=\sin\phi\hat{X}-\cos\phi\hat{Y}$ and $\hat{2}=\mathcal{D}\left[\cos\phi(1+A_1\cos\theta)\hat{X}+\sin\phi(1+A_1\cos\theta)\hat{Y}-A_1\sin\theta\hat{k}\right]$, where $A_1\equiv(\gamma-1)\cos\theta-\gamma\beta$. The coordinate system $\hat{X}\hat{Y}\hat{k}$ is a global orthogonal right-handed system with $\hat{X}$ being the direction along the projection of jet axis on the plane of sky (LWD18; Lan, Wu \& Dai 2016a,b; Toma et al. 2009). In this global coordinate system, the unit velocity vector of the jet element can be expressed as $\hat{\beta}=\sin\theta\cos\phi\hat{X}+\sin\theta\sin\phi\hat{Y}+\cos\theta\hat{k}$. We then transform the $\hat{\overline{e}'}$ into the observer frame through $\vec{\overline{e}}=\mathcal{D}\hat{\overline{e}'}-(\vec{\beta}\cdot\hat{\overline{e}'})(\frac{\gamma^2}{\gamma+1}\vec{\beta}+\gamma\hat{k}')$. We denote $\vec{\overline{e}}=\overline{e}_X\hat{X}+\overline{e}_Y\hat{Y}+\overline{e}_k\hat{k}$. It can be confirmed that $\overline{e}_k=0$. After some calculation, we get $\overline{e}_X=\mathcal{D}\sin(\phi+\chi'_p)$ and $\overline{e}_Y=-\mathcal{D}\cos(\phi+\chi'_p)$. Finally, the local PA in the observer frame can be expressed as
\begin{align}
\chi_p&=\arctan\left(\frac{\overline{e}_Y}{\overline{e}_X}\right)               \nonumber \\
      &=-\arctan\left(\frac{\cos(\phi+\chi'_p)}{\sin(\phi+\chi'_p)}\right)       \nonumber \\
      &=\phi+\chi'_p+\frac{\pi}{2}+n\pi
\end{align}
where n is an integer to make sure that $\chi_p$ is within $[-\pi/2,\pi/2]$. The expression of $\chi'_p$ is
\begin{align}
\chi'_p&=\frac{1}{2}\arctan\left(\frac{\langle U'_p\rangle}{\langle Q'_p\rangle}\right)   \nonumber \\
       &=\frac{1}{2}\arctan\left(\frac{\langle (\sin\theta'_B)^{1-m}\sin(2\phi'_B)\rangle}{\langle (\sin\theta'_B)^{1-m}\cos(2\phi'_B)\rangle}\right)
\end{align}

In the following, we derive the expressions for the $\sin\theta'_B$, $\sin\phi'_B$ and $\cos\phi'_B$. The coordinate components of unit total magnetic field vector $\hat{B}'_s$ in the two coordinate systems $\hat{x}\hat{y}\hat{\beta}$ and $\hat{1}\hat{2}\hat{k}'$ established above are $\hat{B}'_s=B'_x\hat{x}+B'_y\hat{y}+B'_\beta\hat{\beta}$ and $\hat{B}'_s=\sin\theta'_B\cos\phi'_B\hat{1}+\sin\theta'_B\sin\phi'_B\hat{2}+\cos\theta'_B\hat{k}'$. By comparing the coordinate components in these two system, we get
\begin{align}
\cos\theta'_B&=B'_x\sin\theta'+B'_\beta\cos\theta',  \nonumber \\
\sin\theta'_B&=\sqrt{1-(\cos\theta'_B)^2},           \nonumber \\
\cos\phi'_B&=B'_y/\sin\theta'_B,                     \nonumber \\
\sin\phi'_B&=\frac{B'_\beta-\cos\theta'_B\cos\theta'}{\sin\theta'_B\sin\theta'}.
\end{align}

In a smaller region (where the random component has a fixed direction), let $\theta_r$ and $\phi_r$ be the polar and azimuthal angles of the random magnetic field in $\hat{x}\hat{y}\hat{\beta}$ coordinate system. Thus, we have $\hat{B}'_{rnd}=\sin\theta_r\cos\phi_r\hat{x}+\sin\theta_r\sin\phi_r\hat{y}+\cos\theta_r\hat{\beta}$. If we have $\sin\theta'_B$, $\sin\phi'_B$ and $\cos\phi'_B$ as functions of $\theta_r$ and $\phi_r$ and further assume an isotropic distribution of the random magnetic field in a point-like region, then the average over random magnetic field directions can be expressed for example as
\begin{equation}
\langle\sin\theta'_B\rangle=\frac{1}{4\pi}\int^{\pi}_0\sin\theta_rd\theta_r\int^{\pi}_{-\pi}\sin\theta'_Bd\phi_r.
\end{equation}

We assume that there is no conversion between the ordered component and the random component during the afterglow phase, because the magnetization parameter is relatively low at this stage. In a smaller region, the total magnetic field can be expressed as $\vec{B}'_s=\vec{B}'_{ord}+\vec{B}'_{rnd,rs}=B'_{ord}\hat{B}'_{ord}+B'_{rnd}\hat{B}'_{rnd}$. If we assume $B'_{ord}=\xi_BB'_{rnd}$ as in our previous paper (LWD18), then the unit vector of the total magnetic field is $\hat{B}'_s=\vec{B}'_s/B'_s=(\xi_B\hat{B}'_{ord}+\hat{B}'_{rnd})/\eta$, where $\eta\equiv\sqrt{1+\xi^2_B+2\xi_B\hat{B}'_{ord}\cdot\hat{B}'_{rnd}}$ and $B'_s=\eta B'_{rnd}$ is the strength of the total magnetic field in a smaller region.

We denote $\hat{B}'_{ord}=B'_{ord,x}\hat{x}+B'_{ord,y}\hat{y}+B'_{ord,\beta}\hat{\beta}$. Once we have these components of the ordered magnetic field, the components of the total magnetic field ($\hat{B}'_s$) in $\hat{x}\hat{y}\hat{\beta}$ coordinate system will be obtained and the average over magnetic field directions can be done. Three kinds of ordered magnetic field (i.e., toroidal, aligned and radial) are considered in this paper, and the specific expressions of these components are presented as follows.

\subsection{Mixed Magnetic Field with Aligned Ordered Component}
For a magnetar central engine, a striped wind is formed due to magnetic dipole radiation and an aligned ordered magnetic field component would reside in jet (Spruit et al. 2001). $\hat{B}'_{ord}=\hat{B}'_{A}=-J_{A,y}/A_A\hat{x}+J_{A,x}/A_A\hat{y}$ with $J_{A,x}=\sin\delta_a\cos\theta_V\cos\theta\cos\phi+\cos\delta_a\cos\theta\sin\phi+\sin\theta_V\sin\delta_a\sin\theta$, $J_{A,y}=-\sin\delta_a\cos\theta_V\sin\phi+\cos\delta_a\cos\phi$ and the normalization factor $A_A=\sqrt{J_{A,x}^2+J_{A,y}^2}$ (LWD18).

\subsection{Mixed Magnetic Field with Toroidal Ordered Component}
For a black hole central engine, a jet is powered by BZ mechanism and a corresponding ordered magnetic field component in it is very likely to be toroidal (Spruit et al. 2001). $\hat{B}'_{ord}=\hat{B}'_{T}=-J_{T,y}/A_T\hat{x}+J_{T,x}/A_T\hat{y}$ with $J_{T,x}=-\sin\theta_V\cos\theta\cos\phi+\cos\theta_V\sin\theta$, $J_{T,y}=\sin\theta_V\sin\phi$ and the normalization factor $A_T=\sqrt{J_{T,x}^2+J_{T,y}^2}$ (LWD18).

It is hard to prove $U_{\nu,rs}=0$ using the parity of the integrand in the 3D mixed magnetic field in the toroidal ordered component case. The net magnetic field in a point-like region for this case is the same as that of a pure toroidal ordered magnetic field because the random component is isotropic in the 3D space. The Stokes parameter $U_{\nu,rs}$ is zero in such a pure toroidal ordered magnetic field configuration (Toma et al. 2009; Lan, Wu \& Dai 2016a). Therefore, the Stokes parameter $U_{\nu,rs}$ is also zero for the 3D mixed magnetic field with a toroidal ordered component.

Since the toroidal and aligned ordered components are all assumed to be confined in the shock plane, the following formulae are suitable for both cases,
\begin{align}
B'_{x}&=(\xi_BB'_{ord,x}+\sin\theta_r\cos\phi_{r})/\eta,    \nonumber \\
B'_{y}&=(\xi_BB'_{ord,y}+\sin\theta_r\sin\phi_{r})/\eta,     \nonumber \\
B'_{\beta}&=\cos\theta_{r}/\eta.
\end{align}
The dot product of the ordered and random components is $\hat{B}'_{ord}\cdot\hat{B}'_{rnd}=\sin\theta_r(B'_{ord,x}\cos\phi_r+B'_{ord,y}\sin\phi_r)$.

\subsection{Mixed Magnetic Field with a Radial Ordered Component}
A radial ordered component is discussed here as a possible case. Here, the lateral expansion is not considered. We have $\hat{B}'_{ord}=\hat{B}'_{R}=\hat{\beta}$, and
\begin{align}
B'_{x}&=\sin\theta_r\cos\phi_r/\eta,    \nonumber \\
B'_{y}&=\sin\theta_r\sin\phi_r/\eta,    \nonumber \\
B'_{\beta}&=(\xi_B+\cos\theta_r)/\eta.
\end{align}
The dot product in the radial ordered component case is $\hat{B}'_{ord}\cdot\hat{B}'_{rnd}=\cos\theta_r$. It can be proved that $U_{\nu,rs}=0$ in this case, because the integrand is an odd function for the $\phi$ integral.

\section{Polarization of the Jet Emission}
Jet emission includes both the RS and FS contributions. In the FS region, we assume that the magnetic field is random and confined in the shock plane. Polarization calculations for such random magnetic field follow our previous work (Lan, Wu \& Dai 2016a). It should be noted that the local PD $\Pi_p$ is positive which gives the polarization value while the polarization direction is described by the local PA. Therefore, we take $\Pi_p=|\bar{q}'|/\bar{f}'$ with $\bar{q}'=-\pi_0\langle (\sin\theta'_B)^{1-m}\cos(2\phi'_B)\rangle$ and $\bar{f}'=\langle (\sin\theta'_B)^{1-m}\rangle$. If $\bar{q}'$ is larger than 0, then the local PA $\chi_p=\phi+3/2\pi$. If $\bar{q}'$ is smaller than 0, then $\chi_p=\phi$. The Stokes parameter U from the FS region with such a 2D random magnetic field is 0, i.e. $U_{\nu,fs}=0$.

Final PD and PA for the jet emission with an aligned ordered component in its RS region can be expressed as
\begin{equation}
\Pi=\frac{\sqrt{Q_{\nu}^2+U_{\nu}^2}}{F_{\nu}}
\end{equation}
\begin{equation}
\chi=\frac{1}{2}\arctan\left(\frac{U_{\nu}}{Q_{\nu}}\right)
\end{equation}
where $Q_{\nu}=Q_{\nu,rs}+Q_{\nu,fs}$, $U_{\nu}=U_{\nu,rs}$ and $F_{\nu}=F_{\nu,rs}+F_{\nu,fs}$ are the Stokes parameters Q, U and the flux of the jet. $Q_{\nu,fs}$ and $F_{\nu,fs}$ represent the Stokes parameter Q and the flux of the FS region. In this case, PD is always positive and represent the magnitude, while the polarization direction is depicted by its PA\footnote{In order to get the correct PA, the signs of Stokes parameters are also needed (LWD18).}.

Final PD for the jet emission with toroidal or radial ordered component in its RS region is
\begin{equation}
\Pi=\frac{Q_{\nu}}{F_{\nu}}
\end{equation}
Because the Stokes parameter U from the jet is 0, the sign of the PD is determined by the sign of Stoke parameter Q, i.e., PD can be positive or negative. Here, the sign of PD has its meaning, which represent the direction of polarization, i.e., polarization direction of $PD>0$ will have a $90^\circ$ difference with that of $PD<0$\footnote{It should be noted that PD defined in Eq. (2.55) of Rybicki \& Lightman (1979) is always positive. Here it could be negative, depending on the sign of the Stokes parameter $Q$. The sign of PD only represents its polarization direction and the polarization magnitude is described by its absolute value. This description is also seen in Sari (1999) and GK03.}.

\section{Numerical Results}
In our previous paper (LWD18), a 2D ``mixed" magnetic field was considered. As a more general case, a 3D mixed magnetic field is investigated here. In LWD18, the magnetic energy is assumed to be conserved in a smaller region where the direction of the random magnetic field is fixed. We found that the ``random'' component is perpendicular to the ordered component in a smaller region and further assumed that they are perpendicular to each other in a point-like region. The magnetic energy conservation is not necessarily held in a smaller region, so that this condition is not assumed here. Thus there is no constraint on the direction of the random magnetic field. We further assume that the random component is isotropic in a point-like region (i.e., it is 3D isotropic). We find the magnetic energy is conserved in a point-like region (although it is not conserved in a smaller region) because the averaged dot product of the two components is zero ($\langle\hat{B}'_{ord}\cdot\hat{B}'_{rnd}\rangle=0$), and we thus have the strength of the total magnetic field in a point-like region $B'=\sqrt{B^{'2}_{ord}+B^{'2}_{rnd}}$. It should be noticed that although the direction of the random component is stochastic and assumed to be isotropic, its value is a constant at arbitrary burst source time.

The radiation parameters for the RS and FS regions are taken as $\epsilon_{B,rs}=10^{-3}$, $\epsilon_{B,fs}=10^{-5}$, $\epsilon_{e,rs}=\epsilon_{e,fs}=0.1$, and $p_{rs}=p_{fs}=2.5$. The source is assumed to be located at redshift $z=0.1$. We take $\xi_B=1$ when calculating the time evolution of the flux and polarization properties. A flat Universe is adopted with $\Omega_M=0.27$, $\Omega_\Lambda=0.73$ and $H_0=71{\rm \,km\,s^{-1}\,Mpc^{-1}}$.

\subsection{Polarization Properties of the Emission from the RS region}
The dynamics used here is the same as that in the thin shell case of a narrow sub-jet in LWD18. Dynamical parameters we take are $E_{iso}=10^{51}\,{\rm ergs}$, $\gamma_0=200$, $\Delta_{0}=10^{10}\,{\rm cm}$ and $\theta_j=0.03$ rad. $E_{iso}$ is the isotropic equivalent energy of the jet. $\gamma_0$ and $\Delta_0$ are the initial Lorentz factor and initial width of the jet. The redshift of the source is taken as $z=0.1$. We take the number density of the inter-stellar medium as $n_1=1\,{\rm cm}^{-3}$. Three kinds of the ordered magnetic field components are considered: aligned, toroidal and radial, in which the first two kinds are assumed to be confined in the shock plane. These three kinds of ordered component configurations are shown in Fig. 1. Corresponding PD and PA for the RS emission of the aligned ordered component case are expressed as $\Pi_{rs}=\sqrt{Q_{\nu,rs}^2+U_{\nu,rs}^2}/F_{\nu,rs}$ and $\chi_{rs}=\frac{1}{2}\arctan(U_{\nu,rs}/Q_{\nu,rs})$\footnote{The signs of the Stokes parameters $Q_{\nu,rs}$ and $U_{\nu,rs}$ are also needed in ordered to get the real $\chi_{rs}$ (LWD18).}. PDs for the toroidal and radial ordered component cases are $\Pi_{rs}=Q_{\nu,rs}/F_{\nu,rs}$ and PAs in these two cases have a $90^\circ$ difference for $\Pi_{rs}>0$ and $\Pi_{rs}<0$.

The ordered magnetic field component in Fig. 2 is aligned. The viewing angle is taken as $\theta_V=0$. PD is roughly constant ($\sim17\%$) before the RS crossing time and begins to decrease quickly after it. Then PD begins to rise after $\sim200$ s and finally keeps roughly as a constant ($\sim20\%$) after $10^4$ s. The rise of PD at late stage is somewhat hard to understand. PA is roughly a constant during evolution.

The ordered magnetic field component in Fig. 3 is toroidal. The viewing angle is taken as $\theta_V=0.5\theta_j$. If we take $\theta_V=0$ in the toroidal ordered component case, there is no net polarization because of axial symmetry. In this case, the Stokes parameter $U_{\nu,rs}$ is 0. As mentioned above, the final PD from the RS region is $\Pi_{rs}=Q_{\nu,rs}/F_{\nu,rs}$. $\Pi_{rs}$ can be positive or negative, depending on the sign of Stokes parameter $Q_{\nu,rs}$. PA for $\Pi_{rs}>0$ has a $90^\circ$ difference from that of $\Pi_{rs}<0$. PD is roughly $\sim17\%$ before the RS crossing time and begins to decay sharply after it. After the RS crossing time, there is a PD bump for the the toroidal ordered component case while it is a PD valley for the aligned case.

The ordered magnetic field component in Fig. 4 is radial. The viewing angle is also taken as $\theta_V=0.5\theta_j$. In this case, same as in toroidal ordered component case, Stokes parameter $U_{\nu,rs}$ is zero. PD is roughly zero before and slightly after the RS crossing time. Subsequently, with an increase of the $1/\gamma$ cone, there are two PD bumps and PA changes abruptly by $90^\circ$ when PD evolves from positive to negative. After $2\times10^5$ s, PD rises and flux decays very fast. Here the EATS effect is considered and the integration over $\theta$ is on the EATS. One value of $\theta$ corresponds to one jet radius $R$ and $R$ decreases with increasing $\theta$, i.e., a larger $\theta$ value corresponds to a smaller $R$ on an EATS. After the RS crossing time, the flux from the RS region drops to zero after some radius $R_0$ where the observational frequency exceeds the cutoff frequency (Kobayashi 2000; Zou et al. 2005). Therefore, on an EATS, only the flux from larger $\theta$ (smaller $R$) is non-zero at late times. And also with increasing the observational time, the radius corresponding to same $\theta$ value on the EATS also increases. Finally, there is only one $\theta\ (\sim0.0447)$ circle with non-zero flux. So the flux drops sharply and PD rises to some larger value.

The insets of Figs. 2 - 4 show the PD evolution with $\xi_B$ and the black diamonds are our calculation points. They are calculated at the RS crossing time for the aligned and toroidal cases while calculated at 100 s for the radial case because PD of the radial case at early time is roughly 0. When $\xi_B$ approaches zero, i.e., the magnetic field is dominated by the random component and PD of the RS emission approaches zero, no matter what kind of ordered component in the RS region is. When $\xi_B\gg1$, i.e., the ordered component dominates the total field, PDs in the aligned and toroidal cases all approach their value ($\sim60\%$) in corresponding pure ordered magnetic field. For three ordered component cases, PD curve is very steep when $\xi_B\leq\sqrt{2}$ and keeps increase when $\xi_B<10$. When $\xi_B\geq10$, PD is roughly a constant for the aligned and toroidal cases while it goes on increase for the radial case.

\subsection{Polarization Properties of the Emission from the Whole Jet}
In the FS region, the random magnetic field is assumed to be 2D and confined in the shock plane. With the above formulae and parameters, we then calculate the light curves and polarization properties of the FS region numerically. The results are shown in Fig. 5. When $\theta_V=0$, PD is zero. When $\theta_V=0.5\theta_j$, there are two peaks in PD curve and PA changes abruptly by $90^\circ$ when PD evolves from negative to positive.

The light curves and polarization evolution of the emission from the whole jet including both the RS and FS contributions are exhibited in Fig. 6. The RS emissions of the jet are shown in Figs. 2, 3, and 4, while the FS emissions are shown in Fig. 5. Similar to the thin shell case of Lan, Wu \& Dai (2016a) and LWD18, for the aligned and toroidal cases, there is a PD bump during the RS crossing time but the PD value here is smaller than that in the pure ordered magnetic field. PDs for toroidal and radial cases can be smaller than 0 because the total Stokes parameter of the jet, $U_\nu$, is 0. The polarization direction will change abruptly by $90^\circ$ when the PD changes its sign. Here, for comparison, PAs for toroidal and radial cases are also plotted. PA for the aligned case is roughly a constant but changes abruptly by $90^\circ$ during evolution for the toroidal and radial cases.

\section{Application to GRB 170817A}
The recent detection of the gravitational wave (GW) event GW170817 from double neutron star merger opens a new era of multi-messenger astronomy (Abbott et al. 2017).
A weak and short burst GRB 170817A was observed $\sim$ 1.7 seconds after the GW event (Goldstein et al. 2017; Savchenko et al. 2017; Zhang et al. 2018).
Following GRB 170817A a kilonova called AT 2017gfo was observed by Valenti et al. (2017).
As the kilonova gets dim, multi-band afterglows of this burst arise continually to $\sim$ 120 days (Ruan et al. 2018; Lazzati et al. 2017; D'Avanzo et al. 2018; Lyman et al. 2018; Margutti et al. 2018; Alexander et al. 2017; Hallinan et al. 2017; Kim et al. 2017; Mooley et al. 2017),
which is unexpected from a top-hat jet model for GRB afterglows. Recently, Corsi et al. (2018) had reported their upper limit observations ($\sim12\%$) of the radio band PD at 244 days.

Several scenarios have been proposed to interpret the afterglow of GRB 170817A, such as the structured jet (Lazzati et al. 2017; Ruan et al. 2018; Lyman et al. 2018; Troja et al. 2018; Resmi et al. 2018), the cocoon with the specific distribution of $\gamma\beta$ (Mooley et al. 2018b; Nakar \& Piran 2018; Troja et al. 2018; Fraija \& Veres 2018), and the pulsar-powered relativistic jet (Geng et al. 2018). Although the cocoon model for GRB 170817A is inconsistent with the observation of VLBI (Mooley et al. 2018a),
it is still hard to distinguish the structured jet from the pulsar-powered relativistic jet merely through the light curves.
The corresponding polarimetric observation is thus invoked and may help to distinguish different models.
Recently, Gill and Granot (2018) had predicted
the polarization properties within the structured jet model (including both the Gaussian and power-law jets) and the cocoon model under different MFCs for the radio afterglow of GRB 170817A. They found that a large PD ($\sim 60\%$) was expected after the peak of the light curve for the structured jet,
while it was relatively low ($<10\%$) for the energy injection model.

If the merger remnant of GW170817 is a neutron star, energy flow from the magnetic dipole radiation (MDR) of the neutron star will be injected into the
outer jet (Dai \& Lu 1998a; Dai 2004). Initially, the injected wind is dominated by the Poynting flux. During its propagation, some dissipation processes happen, converting the magnetic energy into the $e^+e^-$ pairs and simultaneously disturbing the magnetic field lines. Therefore, a 3D mixed magnetic field is the most likely MFC in the injected wind. When the injected wind catches up and collides with the outer jet, four regions separated by two shocks would be produced (Dai 2004): (1) the unshocked ISM, (2) the forward shocked ISM, (3) the reverse shocked wind and (4) the unshocked wind.

The dynamics of such a system can be solved either analytically (Dai 2004) or numerically (Yu \& Dai 2007; Beloborodov \& Uhm 2006; Uhm 2011; Uhm et al. 2012; Geng et al. 2016). Adopting the method in Beloborodov \& Uhm (2006) to obtain the dynamics and assuming
the main emission mechanism is synchrotron radiation, we then apply our polarization calculation of 3D mixed MFC to GRB 170817A.
The spin-down of the central neutron star can be resort to MDR or GW radiation. However, the MDR case is rejected in Geng et al. (2018). Here we only consider the GW radiation braking mechanism. Relevant parameters are taken to be values same as those in Geng et al. (2018) except for the energy participation factor of electrons in the FS region $\epsilon_{e,fs}$ and energy participation factor of random magnetic field in the RS region $\epsilon_{B,rs}$. They are slightly adjusted to match
the new observational data, i.e., $\epsilon_{e,fs}=0.075$, $\epsilon_{B,rs}=0.025$\footnote{The magnetization of the pulsar wind is also comparable to $\epsilon_{B,rs}$
in our calculation. Although this value ($< 1$) seems to be much smaller than that of the outflow launched initially ($\sim 10^3$), it is still within the range of magnetization
at the large distance after considering the conversion of the magnetic energy to kinetic energy (see \citealt{Porth17} for a review).}
for the GW radiation case.
For the neutron star central engine, we consider the aligned ordered magnetic field component in the RS region with $\delta_a=\pi/6$ and $\xi_B=0.9$.
It is pointed that turbulence would quickly destroy the ordered magnetic component \citep{Deng17} when the magnetization
of the wind is too low. However, the magnetization $\sim 0.01$ adopted in our calculation is still above the lower limit
of this reduction according results in \cite{Deng17}.
We also assume that the local PD from the ordered magnetic field is $\Pi_0=0.6$.

The light curves together with the evolution of PD and PA are shown in Fig. 7 for the GW radiation case. PD is initially zero and rises to a peak with a value of $\sim30\%$ around the 8th day due to the off-axis observation of the FS emission. After the bump, PD roughly keeps as a constant of $\sim10\%$ due to the mixed magnetic field in the RS region. In Section 3.1, we conclude that PD increases with increasing $\xi_B$ when $\xi_B\leq\sqrt{2}$ for the aligned ordered component case. Therefore, a PD upper limit of $12\%$ can give a constraint, $\xi_B\leq0.9$, in the GW case. PA is initially a constant, then evolves gradually, and finally keeps roughly as a constant.

The observations by VLBI show that the distance of the emission region between 75 days and 230 days is roughly $2.7\pm0.3$ mas in the plane of sky at the $1\sigma$ confidence level (Mooley et al. 2018a), corresponding to $(4.7\pm0.5)\times10^{18}$ cm with an observational angle of $20^\circ$. In our calculation, the distance of the shock between day 75 and day 230 is roughly $0.73\times10^{18}$ cm for the GW case. The dynamical distance in our results are about one fifth of the observational value. However, the centroid positions of day 75 and day 230 at the high confidence level $\gtrsim4\sigma$ even coincide with each other. Therefore, our dynamics is basically consistent with the high confidence level distance measurements performed by Mooley et al. (2018a).

\section{Conclusions and Discussion}
In this paper, we have considered a more general 3D mixed magnetic field, including both the ordered and random components, in the RS region of the jet. Adopting the pulsar-powered relativistic jet model (Geng et al. 2018), we have predicted the polarization properties of the radio afterglow of GRB 170817A and constrained the parameter $\xi_B$.

A large-scale ordered magnetic field plays an important role in jet's formation and acceleration. It is possible that there is a large-scale ordered magnetic field remnant in the jet. During the GRB prompt phase, internal shocks, magnetic reconnections or the turbulence may happen and disturb the magnetic field lines, leading to a more random field (Zhang \& Yan 2011; Deng et al. 2016; Troja et al. 2017). Therefore, a mixed magnetic field is a more general case. Depending on different formation mechanisms of the jet, we considered three kinds of the ordered magnetic field components, i.e., aligned, toroidal, and radial (Spruit et al. 2001; Granot 2003). The random component of the mixed magnetic field is assumed to be isotropic in the 3D space at every point-like region.

When $\xi_B<10$, as the ordered magnetic field fraction decreases, PD of the RS emission also decreases for the first two kinds (i.e., aligned and toroidal) of the 3D mixed magnetic field which is consistent with the numerical simulation results (Deng et al. 2016). Generally speaking, PD of the RS emission with a 3D mixed magnetic field of the first two kinds is relatively lower than that of the pure ordered magnetic field cases (see Fig. 4 in LWD18) and the actual value of PD depends on $\xi_B$. When $\xi_B$ is equal to zero, the magnetic field is totally random and PD is almost zero. When $\xi_B\gg1$, the ordered component dominates the total magnetic energy and PD approaches its value in the pure ordered magnetic field.

It should be noted that our radial ordered component case is analogous to an anisotropic random magnetic field in Sari (1999), Gruzinov (1999) and GK03, because in Sari (1999) the anisotropy of the random magnetic field is achieved by multiplying a factor to the parallel component of the magnetic field which is very similar to our radial case. In our setup, the random component is isotropic in the 3D space, corresponding to $\langle B^{'2}_{rnd,x}\rangle=\langle B^{'2}_{rnd,y}\rangle=\langle B^{'2}_{rnd,\beta}\rangle$. We take $\xi_B=1$ and our ordered component is in the radial direction (i.e., $\hat{\beta}$ direction), so we have $(B^{'2}_{ord}+\langle B^{'2}_{rnd,\beta}\rangle)/(\langle B^{'2}_{rnd,x}\rangle+\langle B^{'2}_{rnd,y}\rangle)=2$ (equivalently, $b\equiv2\langle B^{'2}_{\parallel}\rangle/\langle B^{'2}_{\perp}\rangle=4$). Our PD evolution is consistent with that of $b>1$ cases in Fig. 1 of GK03. PDs evolve from positive to negative and the maximum values are all reached at the second PD bump. In this paper, the minimum value of $\xi_B$ is zero, corresponding to $b=1$. Since the values of $b$ increases with $\xi_B$, our model cannot calculate the case of $0\leq b<1$.

For the 2D random magnetic field in the FS region, if the EATS effect is considered, the peak of the PD curve is usually later than $1/\gamma=\theta_V-\theta_j$. For jet emission, when the RS emission dominates the early flux, the profile of PD curves is similar to that of the thin shell case in Lan, Wu \& Dai (2016a) and LWD18 (i.e., there is a PD bump during RS crossing time for the aligned and toroidal cases). The PD of the jet emission during the RS crossing time can range from 0 to $60\%$, depending on the $\xi_B$ value in the RS region, configuration of the ordered component and the flux ratio of the RS region to the FS region.

PA of the 3D mixed magnetic field with an aligned ordered component can evolve gradually and can only change abruptly by $90^\circ$ for the 3D mixed magnetic field with toroidal and radial ordered component because the Stokes parameter $U_\nu$ can be non-zero for the aligned case and is always zero for the toroidal and radial cases. Same as that in Lan, Wu \& Dai (2016a), different evolution patterns of PA in a 3D mixed magnetic field also provides a probe of the ordered part of the magnetic field in the RS region of the jet and thus we can distinguish aligned ordered component from others. In addition, PD evolutions of the toroidal and radial cases are very different for non-zero values of $\xi_B$. There is a PD bump around RS crossing time for the toroidal case (of the thin shell) but PD is almost zero for the radial case. Thus we can distinguish the toroidal component from the radial component. Since the magnetic field configuration of the ordered component in a jet corresponds directly to its central engine (Spruit et al. 2001)\footnote{An aligned ordered component in jet corresponds to a magnetar central engine, and a toroidal ordered component is related to a black hole.}, combining the result of Lan, Wu \& Dai (2016a), we conclude that the central engines of GRBs are distinguishable though polarization evolutions, no matter whether the magnetic field in the RS region is purely ordered or mixed.

Finally, we have applied our polarization calculation within the pulsar-powered relativistic jet model to the radio afterglow of GRB 170817A. Through our fitting, we find the strength ratio of the ordered magnetic field component to the random component can range from 0 to 0.9 for the GW radiation case. Our calculations are also basically consistent with the recent distance measurements (Mooley et al. 2018a).

Actually, polarization properties with a mixed magnetic field were discussed by GK03, where a mixed magnetic field was thought to exist in the ambient medium. The possibility of such an assumption is relatively low. A possible emission region that contains a mixed magnetic field is the RS region, of which the ordered magnetic field component may be carried out from the central engine. The random component in GK03 is the same as that discussed in Sari (1999) and Gruzinov (1999). It is 3D and assumed to be isotropic in the shock plane. The anisotropic of a random magnetic field is described by different magnetic field values in different polar directions. In this paper, the random component of the 3D mixed magnetic field is assumed to be isotropic in the 3D space of a point-like region. With this assumption, the magnetic energy is conserved in a point-like region after averaging over the directions of the random component, although it is not conserved in a smaller region.

In GK03, Stokes parameters from the ordered and random components were calculated separately and the flux was assumed to be proportional to the primary power of the magnetic energy. Here, Stokes parameters from the total magnetic field (including both ordered and random components) is calculated. The index of the flux depending on the magnetic energy density will evolve with time.

\acknowledgements
We thank Siming Liu and Siyao Xu for useful discussions. This work is supported by the National Key Research and Development Program of China (Grant No. 2017YFA0402600) and the National Natural Science Foundation of China (grant Nos. 11573014, 11673068, 11725314, 11433009, and 11833003). X.F.W is also partially supported by the Youth Innovation Promotion Association (2011231), the Key Research Program of Frontier Sciences (QYZDB-SSW-SYS005) and the Strategic Priority Research Program ``Multi-waveband gravitational wave Universe'' (grant No. XDB23040000) of the Chinese Academy of Sciences. M.X.L is supported by the Natural Science Foundation of Jiangsu Province (grant No. BK20171109). J.J.G acknowledges the support by the National Postdoctoral Program for Innovative Talents (grant No. BX201700115) and China Postdoctoral Science Foundation
funded project (grant No. 2017M620199).

\begin{figure}
\begin{center}
\includegraphics[width=1\textwidth,angle=0]{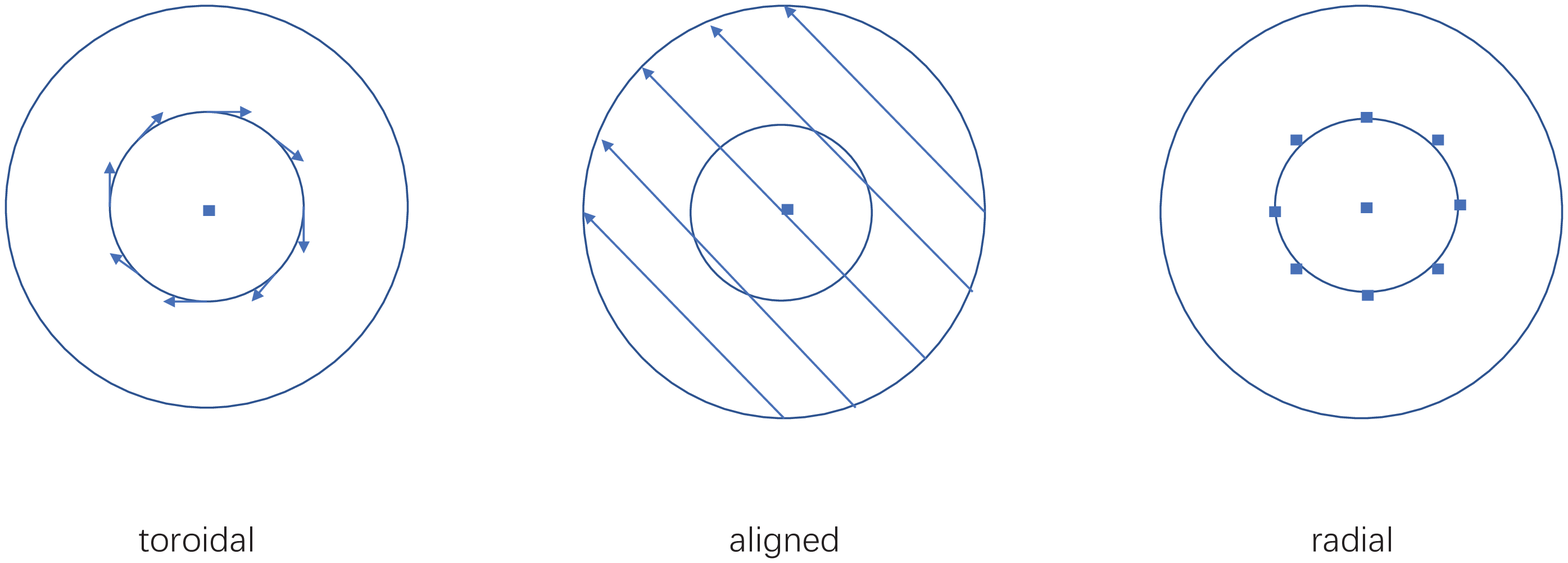}
\caption{Cartoon figure showing the directions of the three different ordered magnetic field configurations considered in this paper. The large circles show the jet region in the shock plane. The arrows show the directions of the ordered magnetic field components in a point-like region. For the toroidal ordered configuration, its direction is along the tangent of the circle centered around the jet axis. For the aligned ordered configuration, its direction is along the parallel lines in jet surface. For the radial ordered configuration, its direction is along the radial direction, which is perpendicular to the shock plane and shown as dots.} \label{fig3}
\end{center}
\end{figure}

\begin{figure}
\begin{center}
\includegraphics[width=1.3\textwidth,angle=0]{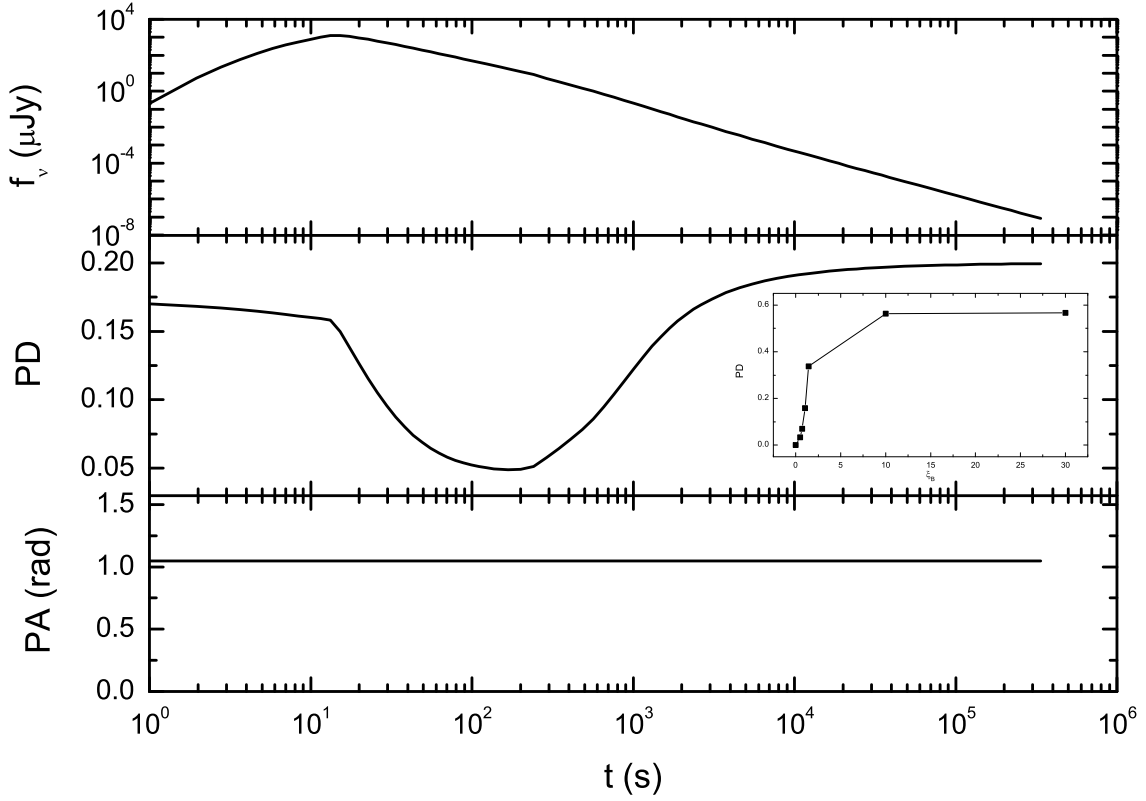}
\caption{Light curve and polarization properties of the RS emission with 3D mixed magnetic field and the ordered magnetic field component is aligned. The upper panel shows the light curve. The medium panel corresponds to PD evolution. The lower panel is for PA evolution. The inset in the mid panel shows the PD evolution with $\xi_B$ values at the RS crossing time. The black diamonds represent our calculation points with $\xi_B=0,\ 1/2,\ 1/\sqrt{2},\ 1,\ \sqrt{2},\ 10$, and 30.} \label{fig1}
\end{center}
\end{figure}

\begin{figure}
\begin{center}
\includegraphics[width=1.3\textwidth,angle=0]{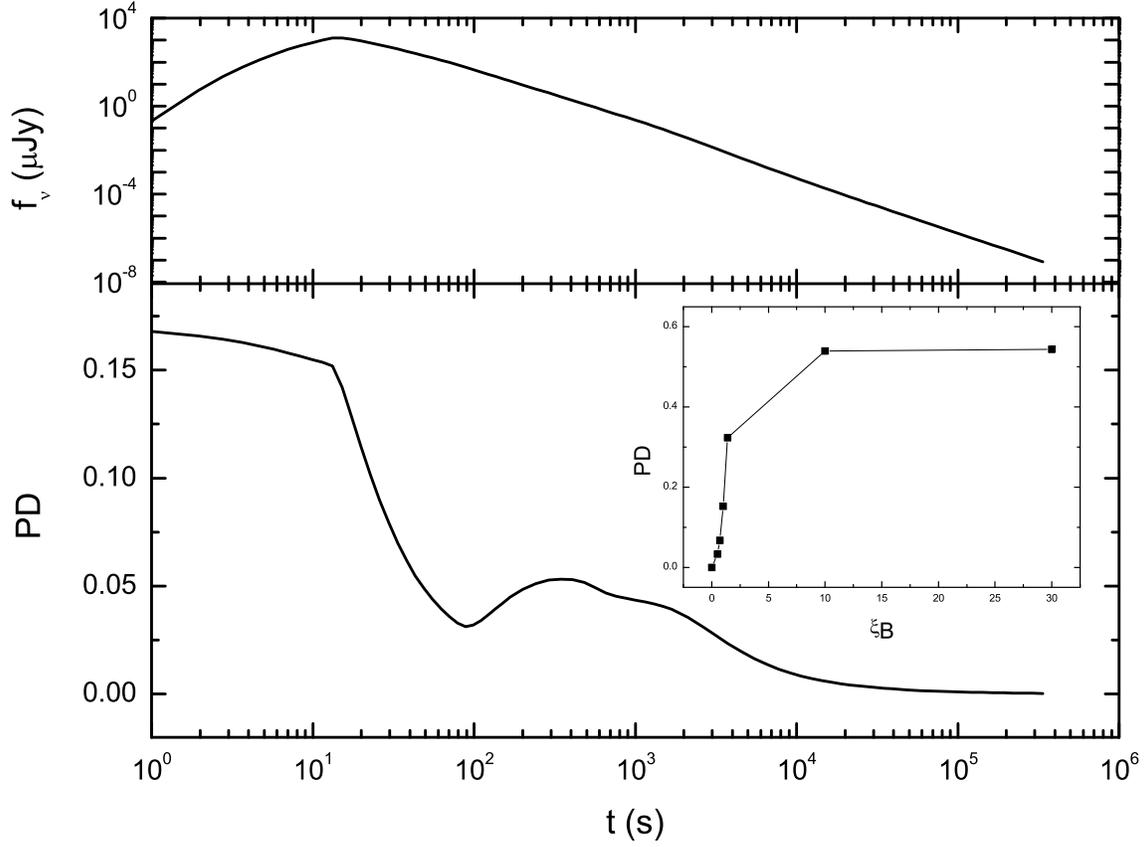}
\caption{Light curve and PD evolution of the RS emission with 3D mixed magnetic field and the ordered component is toroidal. The upper panel shows the light curve. The lower panel corresponds to PD evolution. Same as that in Fig. 2, PD evolution with $\xi_B$ values at the RS crossing time is shown in the inset of the lower panel. PAs in this case have a $90^\circ$ difference between ${\rm PD}>0$ and ${\rm PD}<0$.} \label{fig1}
\end{center}
\end{figure}

\begin{figure}
\begin{center}
\includegraphics[width=1.3\textwidth,angle=0]{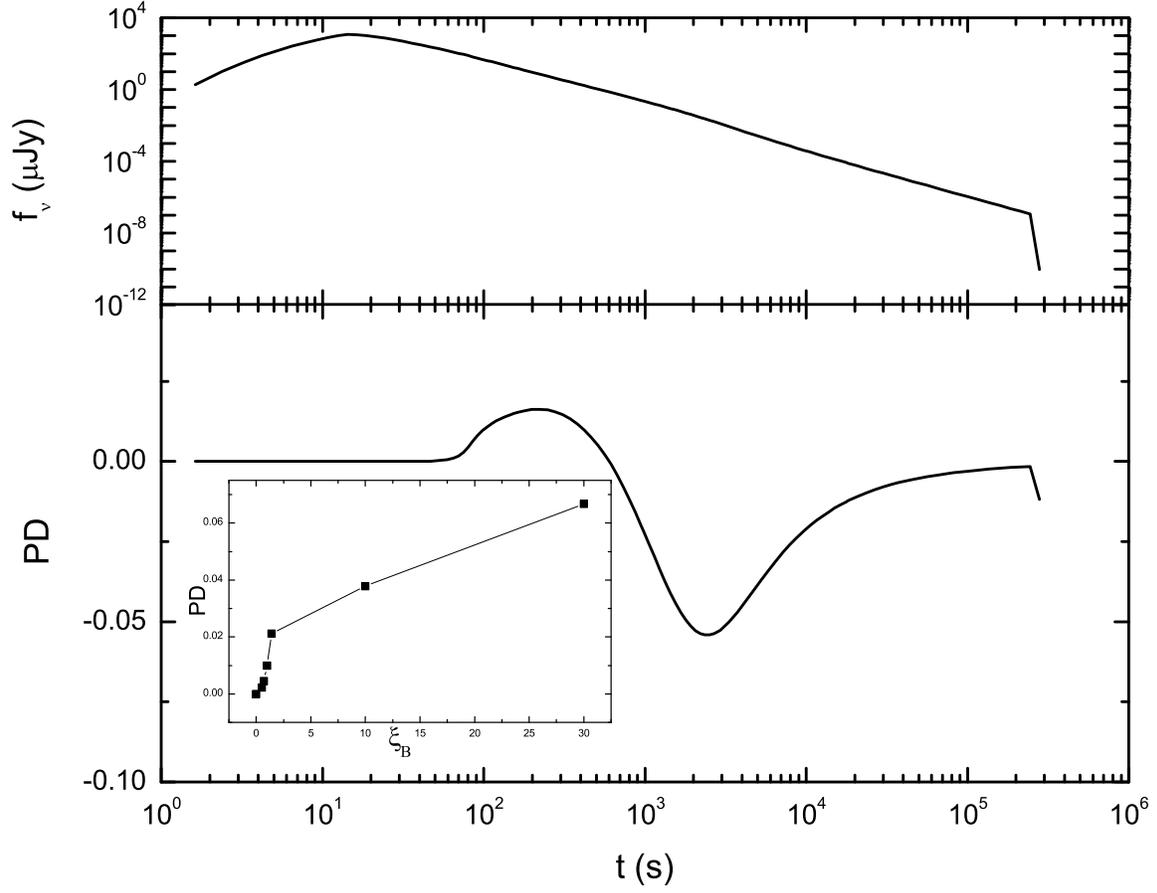}
\caption{Same as Fig. 3, but for the radial ordered component. Since PD is almost zero at early stage in this case, PD evolutions with different $\xi_B$ values are calculated at $t=100$ s.} \label{fig1}
\end{center}
\end{figure}

\begin{figure}
\begin{center}
\includegraphics[width=1.3\textwidth,angle=0]{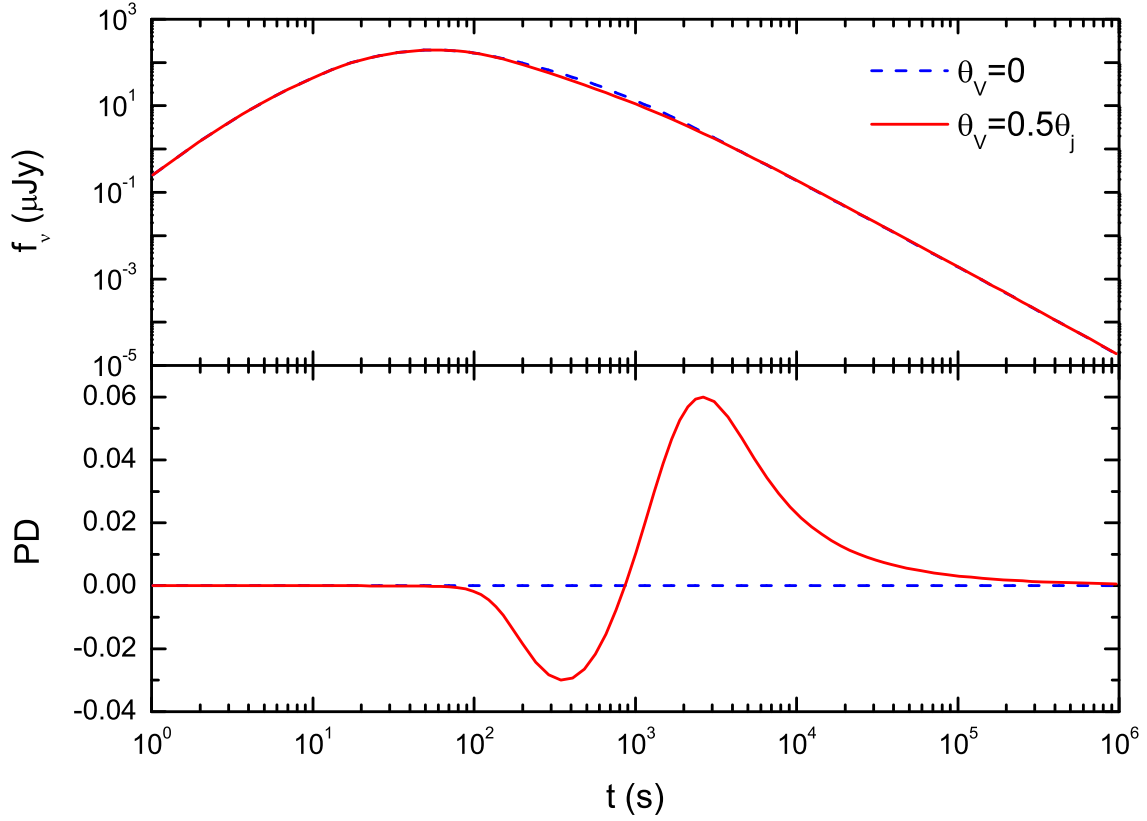}
\caption{Light curves and polarization properties of the FS emission with a 2D random magnetic field confined in the shock plane. The upper panel shows the light curves. The lower panel shows the PD evolutions. The red-solid and blue-dashed lines correspond to FS emission with $\theta_V=0.5\theta_j$ and $\theta_V=0$, respectively. There is a $90^\circ$ PA difference of the FS emission between ${\rm PD}>0$ and ${\rm PD}<0$.} \label{fig2}
\end{center}
\end{figure}

\begin{figure}
\begin{center}
\includegraphics[width=1.0\textwidth,angle=0]{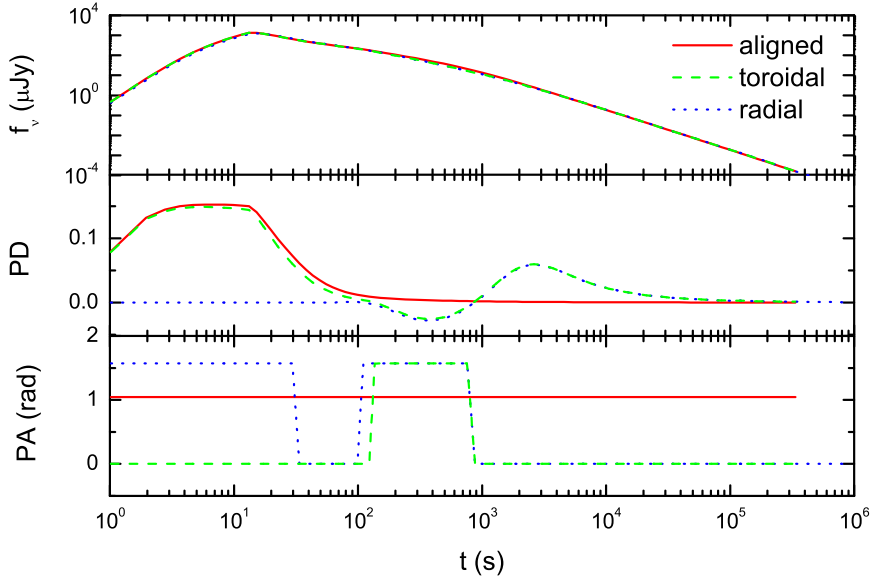}
\caption{Light curves and polarization evolutions of the emission from the whole jet, including the contributions from both the RS and FS regions. The solid, dashed and dotted lines correspond to aligned, toroidal and radial cases, respectively. We take $\theta_V=0$ for the aligned case and $\theta_V=0.5\theta_j$ for the toroidal and radial cases. The RS emissions of the jet are shown in Figs. 2, 3, and 4. FS emissions of the jet are presented in Fig. 5.} \label{fig3}
\end{center}
\end{figure}

\begin{figure}
\begin{center}
\includegraphics[width=1\textwidth,angle=0]{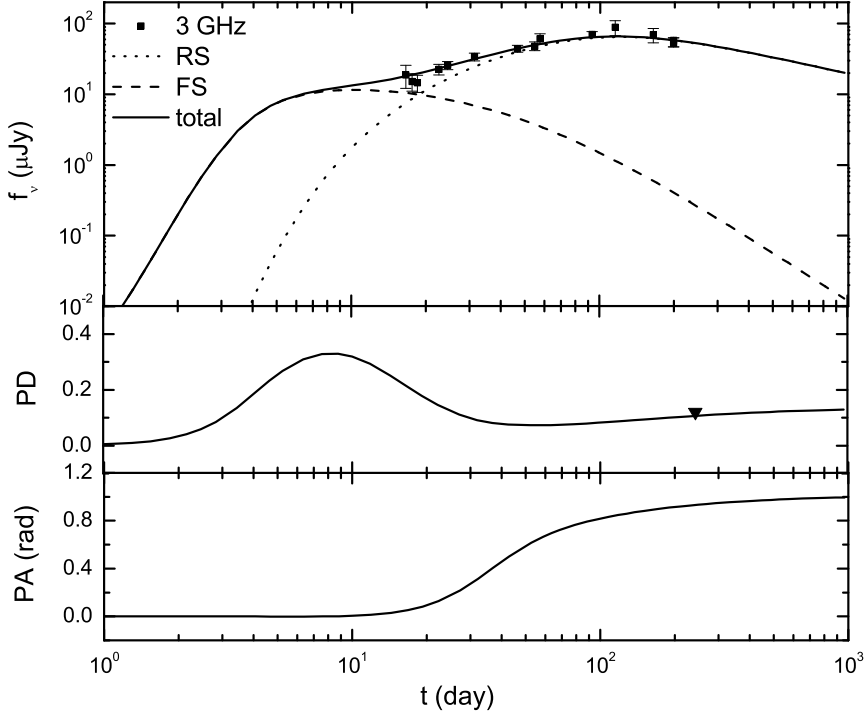}
\caption{Fitting the light curve and polarization evolution of GRB 170817A radio afterglow with the GW radiation braking model. The upper panel shows the light curves. The dotted line corresponds to the RS emission, the dashed line is for the FS emission and the solid line represents the total emission, including both the RS and FS contributions. The medium panel shows the PD evolution. The black triangle shows the upper limit $12\%$ of the PD value at $t=244$ days. The lower panel corresponds to the PA evolution. The radio (3 GHz) data are taken from Alexander et al. (2017), Hallinan et al. (2017), Kim et al. (2017), Mooley et al. (2017), Margutti et al. (2018) and Dobie et al. (2018). The PD upper limit is taken from Corsi et al. (2018).} \label{fig3}
\end{center}
\end{figure}

\end{document}